# Enhanced Multiple Routing Configurations For Fast IP Network Recovery From Multiple Failures

**T. Anji Kumar**  *anji5678@gmail.com*
Dept. of IT/ UCEV
JNTUK
Vizianagaram, 535003, India

**Dr MHM Krishna Prasad** *krishnaprasad.mhm@gmail.com*
Dept. of IT/ UCEV
JNTUK
Vizianagaram, 535003, India

### Abstract

Now a days, Internet plays a major role in our day to day activities e.g., for online transactions, online shopping, and other network related applications. Internet suffers from slow convergence of routing protocols after a network failure which becomes a growing problem. Multiple Routing Configurations [MRC] recovers network from single node/link failures, but does not support network from multiple node/link failures. In this paper, we propose Enhanced MRC [EMRC], to support multiple node/link failures during data transmission in IP networks without frequent global re-convergence. By recovering these failures, data transmission in network will become fast.

## 1. INTRODUCTION

The demand on the Internet has been increased by transforming it from a special purpose network to a common platform for many online services such as online transactions, entertainment and for other e-commerce applications. Internet suffers from slow convergence of routing protocols after a network failure. The central goal in the Internet is the ability to recover from failures [1]. Generally in IP networks, when a node/link failure occurs, the IGP routing protocols like OSPF are used to update the forwarding information based on the changed topology and the updated information is distributed to all routers in the network domain and each router individually calculates new valid routing tables.

The IGP convergence process is slow, as it is reactive i.e., it reacts to a failure after it has happened, and global i.e., it involves all the routers in the domain. This global IP re-convergence is a time consuming process, and a link/node failure is followed by a period of routing instability which results in packet drop. This phenomenon has been studied in both IGP [2] and BGP context [3], and has an adverse effect on real-time applications [4]. Though the different steps of the convergence of IP routing, i.e., detection, dissemination of information and shortest path calculation has been optimized, the convergence time is still too large for applications with real time demands [5]. Since most network failures are short lived [6], too rapid triggering of the re-convergence process can cause route flapping.

Multiple Routing Configurations [MRC] [7] is a proactive and local protection mechanism that allows fast recovery. When a failure is detected, MRC forwards the packets over pre-configured alternative next-hops immediately. Since no global re-routing is performed, fast failure detection mechanisms like fast hellos or hardware alerts can be used to trigger MRC without compromising network stability [8].The shifting of recovered traffic to the alternative link may lead to congestion and packet loss in parts of the network [9]. Ideally, a proactive recovery scheme should not only guarantee connectivity after a failure, but also do so in a manner that does not cause an

T Anji Kumar & Dr MHM Krishna Prasad

unacceptable load distribution. This requirement has been noted as being one of the principal challenges for pre-calculated IP recovery schemes [10].
MRC is a proactive routing mechanism, and it improves the fastness of the routing but it does not protect network from multiple failures. It can protect only from the single link/node failures. Hence, in this paper, using the time slot mechanism, we propose Enhanced Multiple Routing Configurations [EMRC] for fast multiple nodes/links failure recovery.

## 2. ENHANCED MULTIPLE ROUTING CONFIGURATIONS

### 2.1 Motivation
Even though the MRC provides an elegant and powerful hybrid routing framework, it doesn't protect the network from multiple failures and MRC is expensive as it requires more number of backup configurations. Hence, EMRC is designed to support multiple failures by utilizing time slot mechanism and less number of backup configurations.

### 2.2 Basic idea of EMRC
The basic idea of EMRC is as follows: Each source to destination transmission maintains original route. First shortest path is taken as an original route. These shortest paths are calculated by using the OSPF algorithm. Initially, data packets will be transmitted using this original route.

In this source to destination transmission, any sudden occurrence of node or link failure happens, total transmission is collapsed. At this time EMRC uses the timeslot mechanism. If a failure is occurred we will give the timeslot, means give some time to failure recovery before changing the route. Within the timeslot, if the failure is recovered then data is transmitted by using the original route only and if the failure is not recovered, then the data is transmitted by using the backup route and send the probing for failure recovery. During the backup route transmission, if failure is recovered, then backup route transmission is stopped and again reuses the original route. By reusing the original route we can improve the fastness of routing, since the backup route is longer than the original route.

### 2.3 EMRC Approach
EMRC is a threefold approach. First, a set of backup configurations are created, such that every network component is excluded from packet forwarding in one configuration. Second, for each configuration, a routing algorithm like OSPF is used to calculate configuration specific shortest paths and create forwarding tables in each router. Third, a forwarding process is designed which uses the backup configurations to provide fast recovery from a component failure.

### 2.4 Generating Backup Configurations
For generating backup configurations, we adopt an algorithm proposed by Hansen [7]. Our algorithm takes as input the directed graph $G$ and the number $n$ of backup configurations that is intended created. The algorithm will typically be run once at the initial start-up of the network, and each time a node or link is permanently added or removed. We use the notation shown in TABLE1. EMRC configurations are defined by the network topology, which is the same in all configurations, and the associated link weights, which differ among configurations. We formally represent the network topology as a graph $G = (N, A)$, with a set of nodes $N$ and a set of links $A$. In order to guarantee single-fault tolerance, the topology graph $G$ must be bi-connected. A configuration is defined by this topology graph and the associated link weight function:
Definition: A configuration $C_i$ is an ordered pair $(G, W_i)$ of the graph $G$ and a function $W_i : A \rightarrow \{1, \ldots, W_{max}, W_r, \infty\}$ that assigns an integer weight $W_i(a)$ to each link $a \in A$.

We distinguish between the normal configuration $C_0$ and the backup configurations $C_i$, $i > 0$. In the normal configuration $C_0$, all links have "normal" weights $W_0(a) \in \{1, \ldots, W_{max}\}$. We assume that $C_0$ is given with finite integer weights. EMRC is agnostic to the setting of the link weights in $C_0$. In the backup configurations, selected links and nodes must not carry any transit traffic. Still, traffic must be able to depart from and reach all operative nodes. These traffic regulations are imposed by assigning high weights to some links in the backup configurations.

Definition: A link $a \in A$ is isolated in $C_i$ if $W_i(a) = \infty$.
Definition: A link $a \in A$ is restricted in $C_i$ if $W_i(a) = W_r$.

| | |
|---|---|
| $G = (N,A)$ | Graph with set of nodes $N$ and set of links $A$ |
| $C_i$ | The graph having link weights as in configuration $i$ |
| $S_i$ | The set of isolated nodes in configuration $C_i$ |
| $B_i$ | The backbone in configuration $C_i$ |
| $A(u)$ | The set of links from node $u$ |
| $(u,v)$ | The directed link from node $u$ to node $v$ |
| $P_i(u,v)$ | A given shortest path between nodes $u$ and $v$ in $C_i$ |
| $N(p)$ | The nodes on path $p$ |
| $A(p)$ | The links on path $p$ |
| $W_i(u,v)$ | The weight of the link $(u,v)$ in configuration $C_i$ |
| $W_i(p)$ | The total weight of the links in path $p$ in configuration $C_i$ |
| $W_r$ | The weight of a restricted link |
| $n$ | The number of backup configurations to be generated |

**TABLE 1:** Notation

Isolated links do not carry any traffic. Restricted links are used to isolate nodes from traffic forwarding. The restricted link weight $W_r$ must be set to a sufficiently high, finite value to achieve that. Nodes are isolated by assigning at least the restricted link weight to all their attached links. For a node to be reachable, we cannot isolate all links attached to the node in the same configuration. More than one node may be isolated in a configuration. The set of isolated nodes in $C_i$ is denoted $S_i$, and the set of normal (non-isolated) nodes $S_n = N \setminus S_i$.

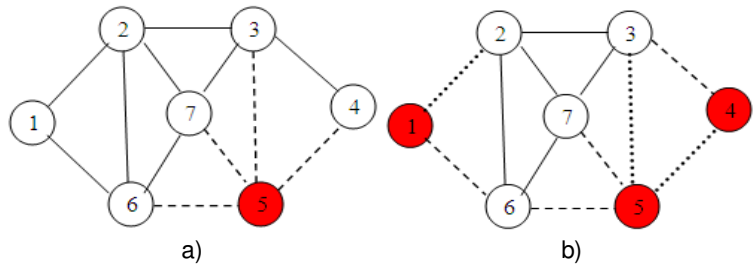

a)                     b)

**FIGURE 1:** Examples of isolated links and isolated nodes.

The purpose of the restricted links is to isolate a node from routing in a specific backup configuration $C_i$, such as node 5 in FIGURE 1.a. In many topologies, more than a single node can be isolated simultaneously. In the example in FIGURE 1.b. three nodes and three links are isolated. Restricted and isolated links are always given the same weight in both directions. EMRC guarantees single-fault tolerance by isolating each link and node in exactly one backup configuration. In each configuration, all node pairs must be connected by a finite cost path that does not pass through an isolated node or an isolated link. A configuration that satisfies this requirement is called valid.

Termination: The algorithm runs through all nodes trying to make them isolated in one of the backup configurations and will always terminate with or without success. If a node cannot be isolated in any of the configurations, the algorithm terminates without success. However, the algorithm is designed so that any bi-connected topology will result in a successful termination, if the number of configurations allowed is sufficiently high.

Complexity: The complexity of the proposed algorithm is determined by the loops and the complexity of the connected method. This method performs a procedure similar to determining whether a node is an articulation point in a graph, bound to worst case $O(|N|+|A|)$. Additionally, for each node, we run through all adjacent links, whose number has an upper bound in the maximum node degree $\Delta$. In the worst case, we must run through all n configurations to find a

T Anji Kumar & Dr MHM Krishna Prasad

configuration where a node can be isolated. The worst case running time for the complete algorithm is then bound by $O(n\Delta|N||A|)$.

### 2.5 Forwarding Procedure for EMRC

When we want to transmit any data from source to destination in the network, first we identify the source node and destination node, after that we look at the shortest path in between them in the original routing table and the data packets are transmitted by using that shortest route. When a data packet reaches a point of failure, the node adjacent to the failure, called the detecting node stops the transmission. At that time, the detecting node gives the timeslot to failure recovery before shifting to the backup route. Within the timeslot, if the failure is recovered then data is transmitted by using the original route only and if the failure is not recovered, then the detecting node is responsible for finding a backup configuration where the failed component is isolated. The detecting node marks the packet as belonging to this configuration, and forwards the packet. From the packet marking, all transit routers identify the packet with the selected backup configuration, and forward it to the egress node avoiding the failed component. Packet marking is most easily done by using specific values in the DSCP field in the IP header. If this is not possible, other packet marking strategies like IPv6 extension headers or using a private address space and tunneling [11] could be used. During the backup route transmission, the detecting node sends the probing signals for failure recovery and if failure is recovered, then backup route transmission is stopped and the data packets are transmitted by reusing the original route. By reusing the original route we can improve the fastness of routing, since the backup route is longer than the original route. If a failure lasts for more than a specified time interval, a normal re-convergence will be triggered. EMRC does not interfere with this convergence process, or make it longer than normal. However, EMRC gives continuous packet forwarding during the convergence, and hence makes it easier to use mechanisms that prevents micro-loops during convergence, at the cost of longer convergence times [12]. If a failure is deemed permanent, new configurations must be generated based on the altered topology.

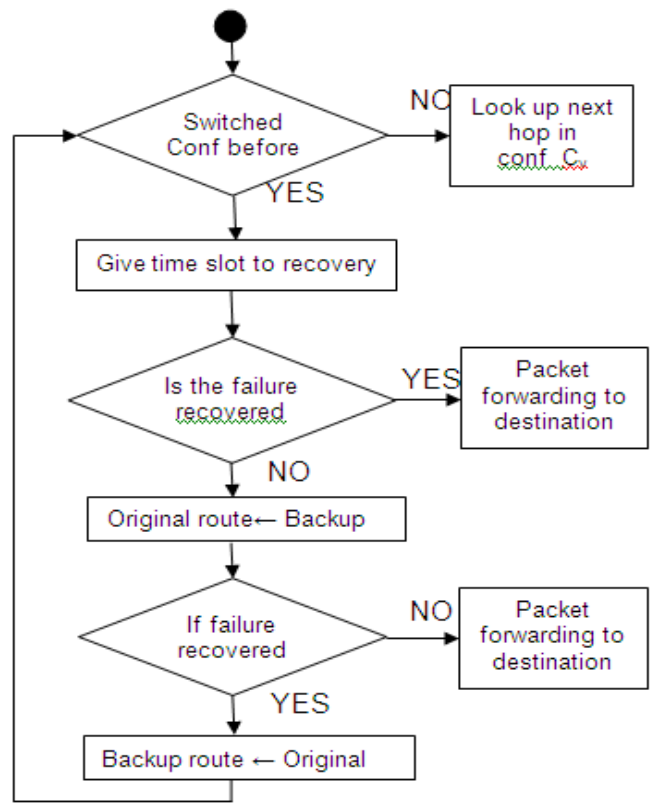

T Anji Kumar & Dr MHM Krishna Prasad

**FIGURE 2:** Packet forwarding state diagram.

## 2.6 Comparison of MRC and EMRC

EMRC is developed from MRC. So, all the processes in EMRC such as backup route finding, shortest path finding and forwarding is same as the MRC. These all are explained and compared by using the following in FIGURE 3 and FIGURE 4.

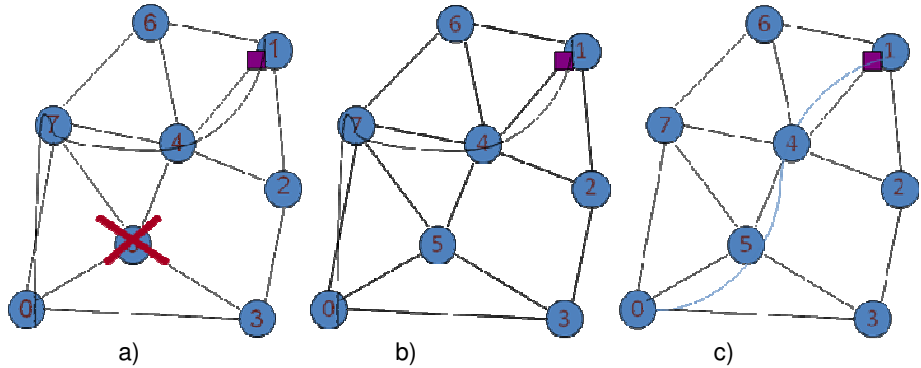

          a)                                         b)                                       c)

**FIGURE 3:** Selection of routes in MRC and EMRC a) At the time of failure occurrence in MRC and EMRC b) After the failure recovery in MRC c) After the failure recovery in EMRC.

As shown in FIGURE 3, we want to transmit the data from node 1 to node 0 by using the shortest path. Hence, in FIGURE 3 the source node is 1 and destination node is 0 and the shortest route is 1-4-5-0. 1-4-5-0 route is taken as a original route. Another route i.e. 1-4-7-0 is a backup configuration, where the node 5 is isolated. In original route, at middle of the transmission, any sudden occurrence of failure of node 5, data transmission is stopped at node 4. At that time MRC selects the backup route i.e. 1-4-7-0 and transmit the data to destination. By using the backup route, total transmission time increases and fastness of the routing decreases.

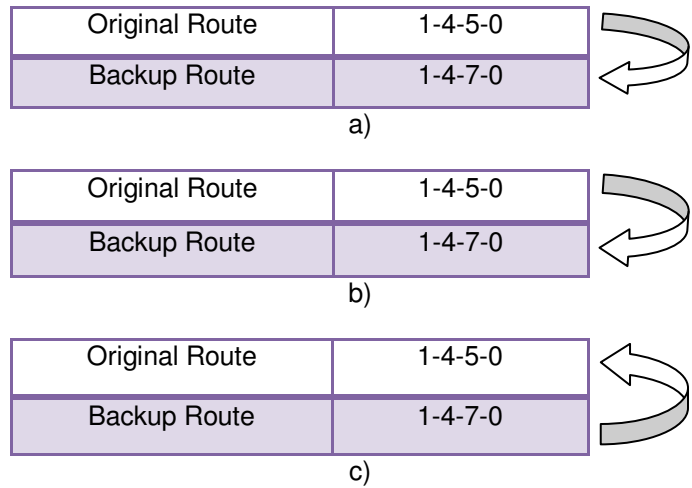

| Original Route | 1-4-5-0 |
|---|---|
| Backup Route | 1-4-7-0 |

a)

| Original Route | 1-4-5-0 |
|---|---|
| Backup Route | 1-4-7-0 |

b)

| Original Route | 1-4-5-0 |
|---|---|
| Backup Route | 1-4-7-0 |

c)

**FIGURE 4:** Selection of routes in MRC and EMRC a) At the time of Failure occurrence in MRC and EMRC b) After the failure recovery in MRC c) After the failure recovery in EMRC.

Using EMRC, at the time of failure of node 5, it gives the timeslot for failure recovery before shifting to the backup route i.e. 1-4-7-0. Within the timeslot if the failure is recovered, then the data is transmitted by using the original route i.e. 1-4-5-0 only. If the failure is not recovered, then the transmission is shifted to the backup route i.e. 1-4-7-0. During the time of backup route transmission we send the probes for failure recovery to the node 5. If at any time, failure is

T Anji Kumar & Dr MHM Krishna Prasad

recovered, we again reuse the original route i.e. 1-4-5-0 and backup route transmission is stopped. In EMRC, by using the timeslot and reusing mechanism, we can improve both i.e. fastness of routing and as well as data transmission.

## 3. EXPERIMENTAL WORK

The Enhanced Multiple Routing Configurations ( EMRC ) scheme is implemented in C++ and TCL ( Tool Command Language ) by using "ns-allinone-2.30" tool and the whole experiment is carried out on the LINUX operating system.

The following FIGURE 5.1 shows the process of simulation in which we have to observe the simulation time as well as the nodes advertising themselves.

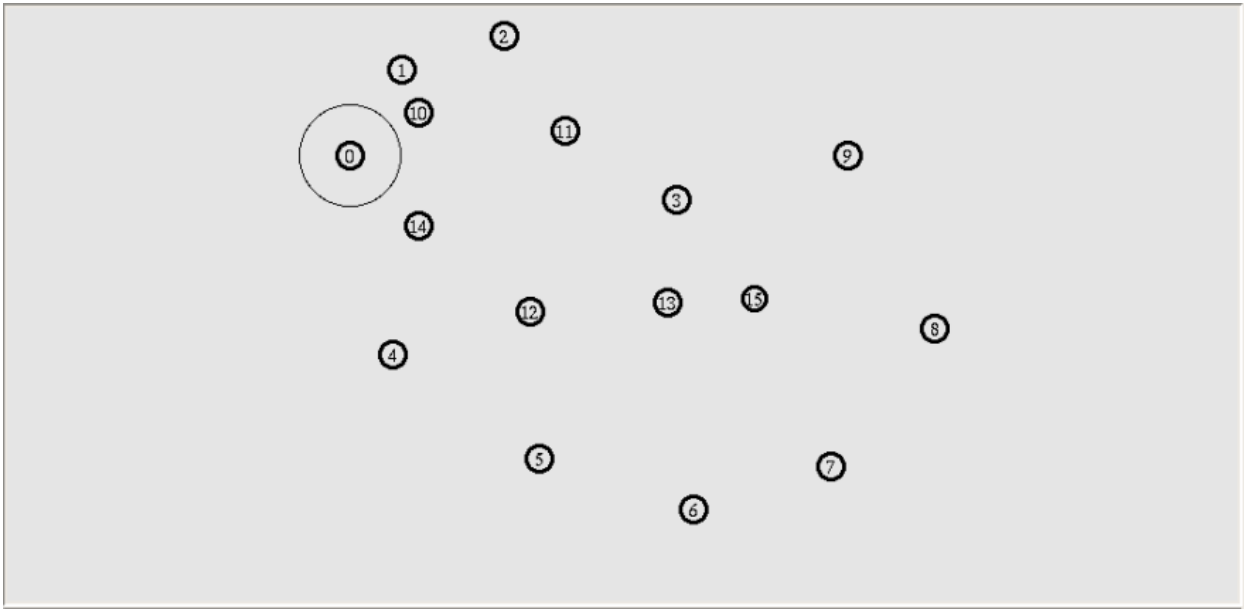

**FIGURE 5.1:** Screenshot of nodes advertising themselves.

The following FIGURE 5.2 shows the route discovery process to the destination. In this process every node tries to know the nearest neighbours for finding the shortest route.

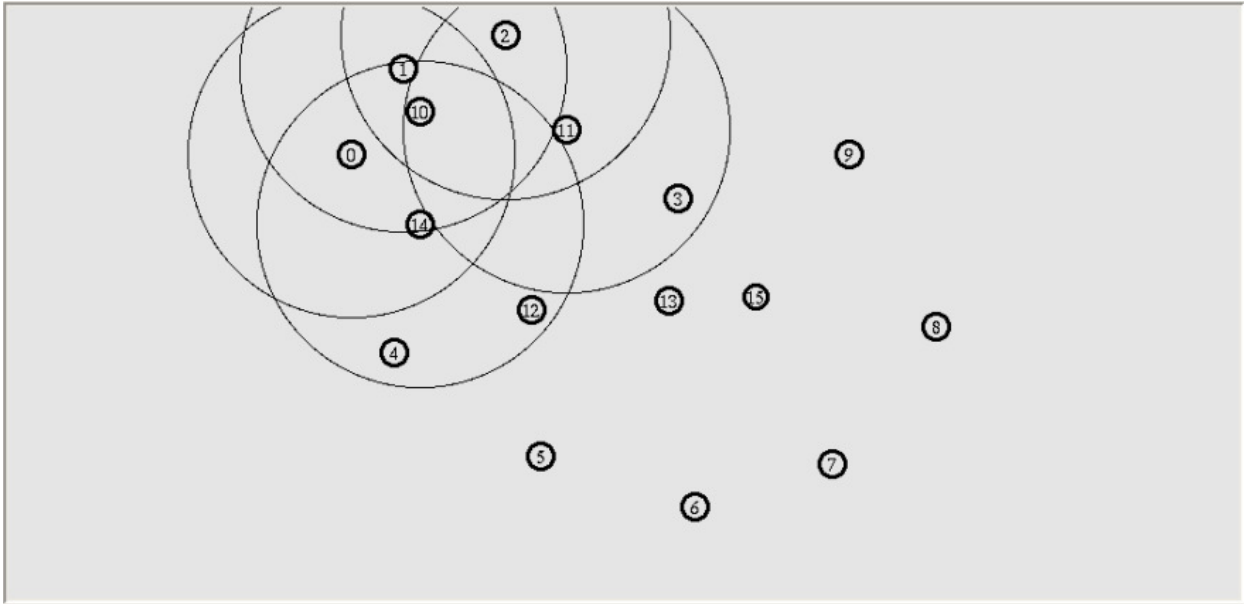

**FIGURE 5.2:** Screenshot of route discovery process.

The following FIGURE 5.3 represents the process of sending acknowledgements to senders after receiving request or data i.e., node 6 send acknowledgements to node 5 and node 13.

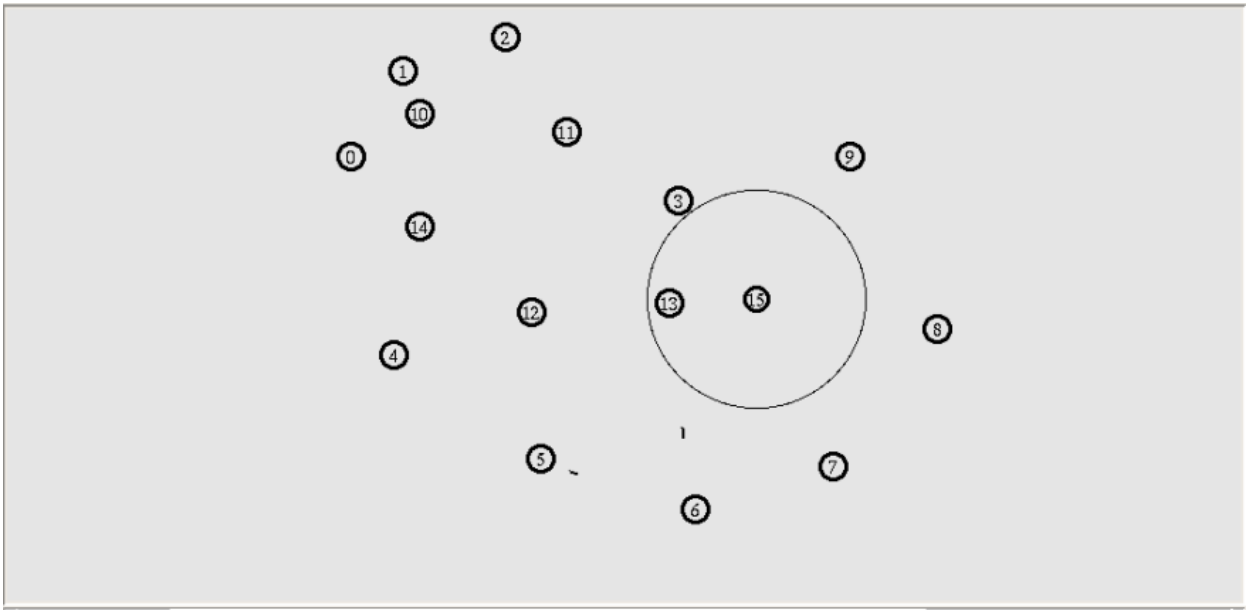

**FIGURE 5.3:** Destination node sends acknowledgements to the best routes.

The following FIGURE 5.4 represents the node 4 receiving the packet from node 14 and sending the acknowledgement to node 14. After that node 4 send data to node 5.

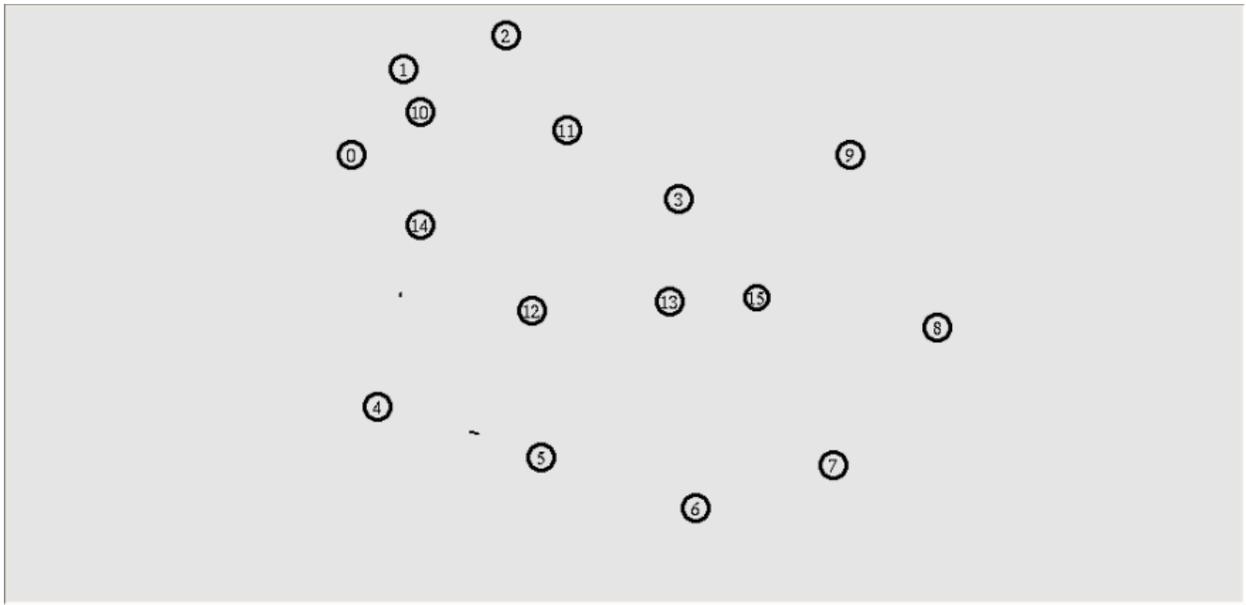

**FIGURE 5.4:** Receives the packet and send acknowledgement to the source.

From the following FIGURE 5.5, one can observe that the node 4 is out of range from node 14 i.e. node 4 is a failure node. So node 14 gives timeslot for failure recovery.

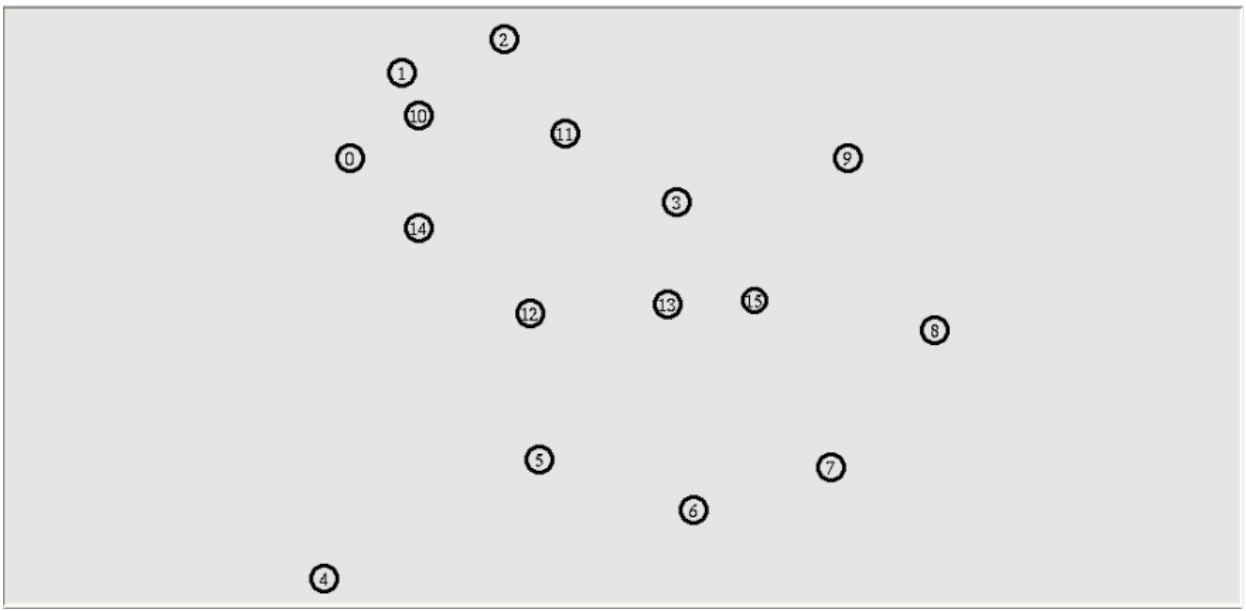

**FIGURE 5.5:** Movement of node 4 out of range of node14, breakage of best route.

From the following FIGURE 5.6, one can observe that within the timeslot, node 4 is not recovered from failure. So, the node 14 transmits the data by using the backup route and probes for node 4 from failure recovery.

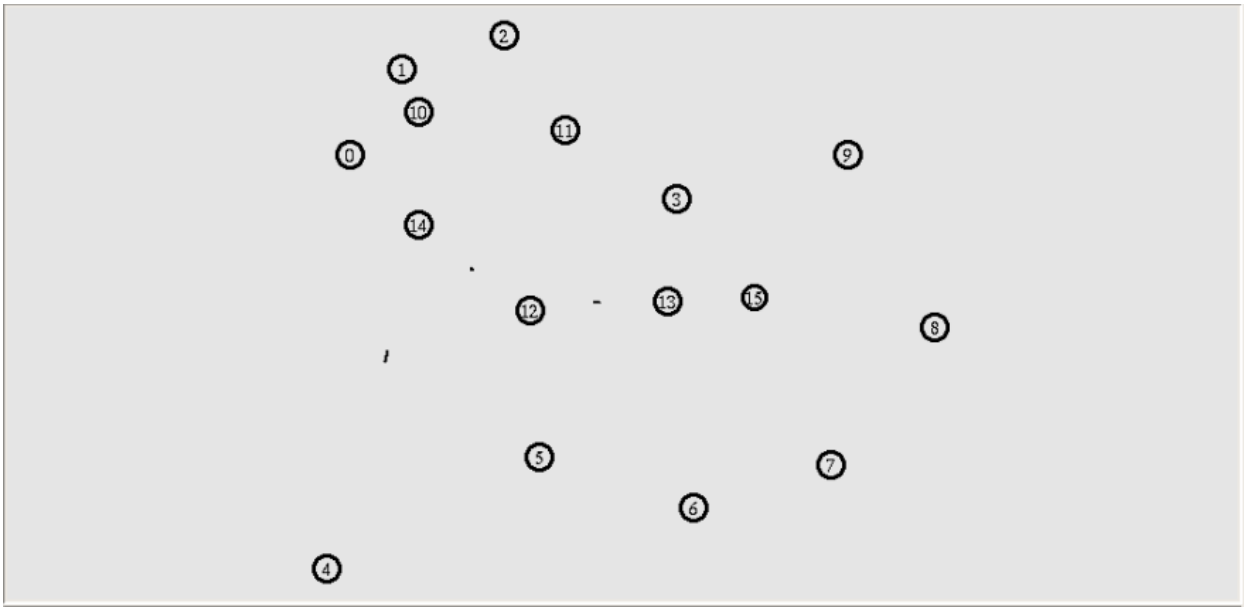

**FIGURE 5.6:** Selection of backup route and probing for node 4.

The following FIGURE 5.7 shows the selection of backup route in the trace file.

```
break
Node-14 =  i didnt receive ack
Main Route Failure
sending packet through Backup Route

Sending through backup route
Route ::
10
14
12
13
15
7
6
Route ::
10
14
12
13
15
7
6
s -t 3.914209689 -Hs 14 -Hd 12 -Ni 14 -Nx 120.00 -Ny 420.00 -Nz 0.00 -Ne -1.000000 -Nl RTR -Nw ---
-Ma 0 -Md 0 -Ms 0 -Mt 0 -Is 14.255 -Id 12.255 -It amrc -Il 1044 -If 0 -Ii 52 -Iv 32

Probing for node 4

0
r -t 3.923975329 -Hs 12 -Hd 12 -Ni 12 -Nx 250.00 -Ny 320.00 -Nz 0.00 -Ne -1.000000 -Nl RTR -Nw ---
```

**FIGURE 5.7:** Selection of backup route and probing for node 4, see from the trace file.

The following FIGURE 5.8 shows that after the recovery of node 4 from failure, it sends the reply to the node14 for the probing signal.

T Anji Kumar & Dr MHM Krishna Prasad

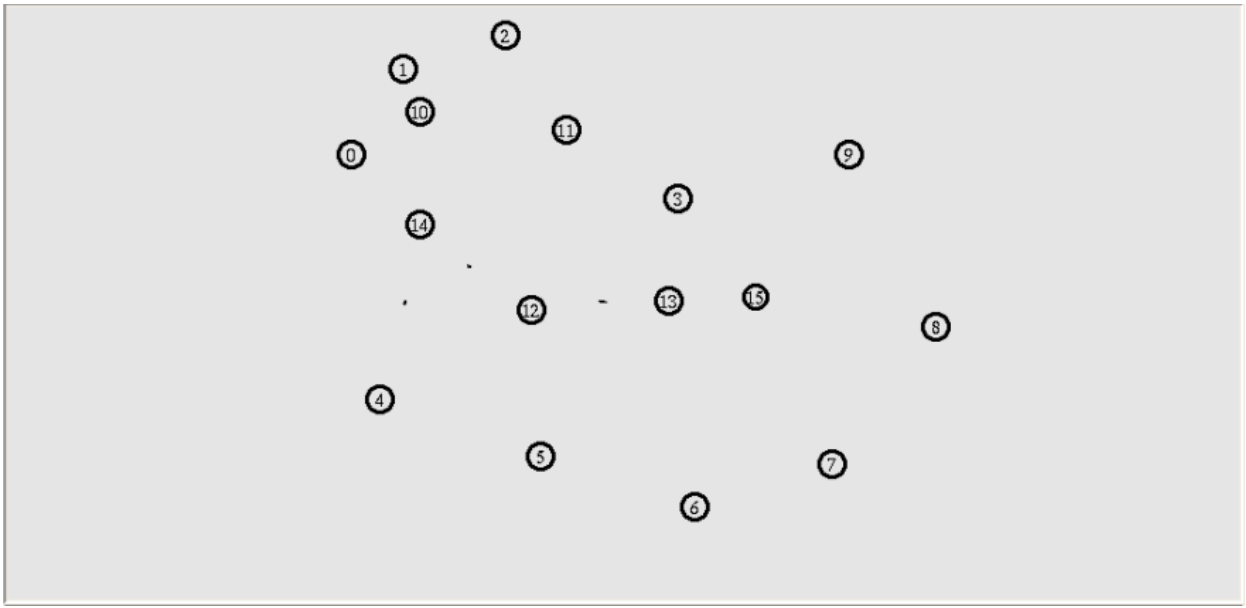

**FIGURE 5.8:** Sending the reply to node 14 for the probing signal after the node 4 is recovered from failure.

The following FIGURE 5.9 shows the use of recovered original route after the node 4 is recovered from failure.

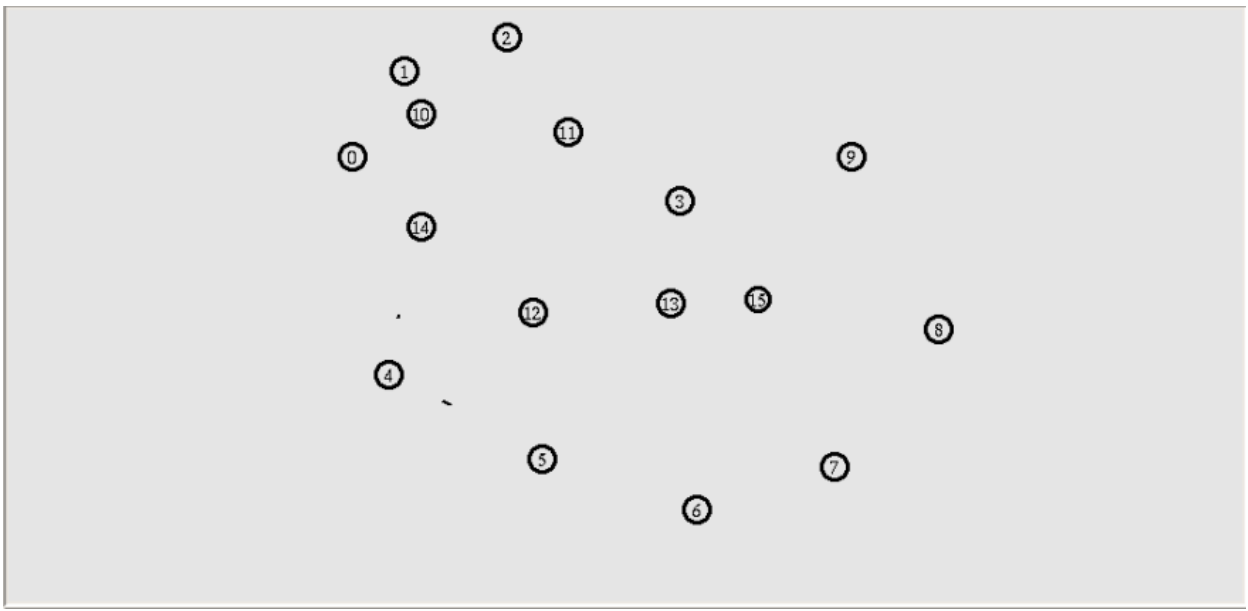

**FIGURE 5.9:** Reconstruction of the original route.

The following FIGURE 5.10 shows the reconstruction of original route after the node 4 failure recovery in the trace file.

```
Node-4 == Sending Ack back to node 14
s -t 4.942351534 -Hs 4 -Hd 14 -Ni 4 -Nx 90.00 -Ny 270.00 -Nz 0.00 -Ne -1.000000 -Nl RTR -Nw --- -
Ma 0 -Md 0 -Ms 0 -Mt 0 -Is 4.255 -Id 14.255 -It amrc -Il 532 -If 0 -Ii 85 -Iv 32
s -t 4.942351534 -Hs 4 -Hd 5 -Ni 4 -Nx 90.00 -Ny 270.00 -Nz 0.00 -Ne -1.000000 -Nl RTR -Nw --- -Ma
13a -Md 4 -Ms e -Mt 800 -Is 4.255 -Id 5.255 -It amrc -Il 1044 -If 0 -Ii 81 -Iv 30
r -t 4.948165063 -Hs 14 -Hd 14 -Ni 14 -Nx 120.00 -Ny 420.00 -Nz 0.00 -Ne -1.000000 -Nl RTR -Nw ---
-Ma 13a -Md e -Ms 4 -Mt 800 -Is 4.255 -Id 14.255 -It amrc -Il 532 -If 0 -Ii 84 -Iv 32

Node-14 == Received Ack

r -t 4.958095654 -Hs 5 -Hd 5 -Ni 5 -Nx 260.00 -Ny 150.00 -Nz 0.00 -Ne -1.000000 -Nl RTR -Nw --- -
Ma 13a -Md 5 -Ms 4 -Mt 800 -Is 4.255 -Id 5.255 -It amrc -Il 1044 -If 0 -Ii 80 -Iv 30

Node-5 == Sending Ack back to node 4
s -t 4.958095654 -Hs 5 -Hd 4 -Ni 5 -Nx 260.00 -Ny 150.00 -Nz 0.00 -Ne -1.000000 -Nl RTR -Nw --- -
Ma 0 -Md 0 -Ms 0 -Mt 0 -Is 5.255 -Id 4.255 -It amrc -Il 532 -If 0 -Ii 86 -Iv 32
s -t 4.958095654 -Hs 5 -Hd 6 -Ni 5 -Nx 260.00 -Ny 150.00 -Nz 0.00 -Ne -1.000000 -Nl RTR -Nw --- -
Ma 13a -Md 5 -Ms 4 -Mt 800 -Is 5.255 -Id 6.255 -It amrc -Il 1044 -If 0 -Ii 80 -Iv 29
r -t 4.964149735 -Hs 4 -Hd 4 -Ni 4 -Nx 90.00 -Ny 270.00 -Nz 0.00 -Ne -1.000000 -Nl RTR -Nw --- -Ma
13a -Md 4 -Ms 5 -Mt 800 -Is 5.255 -Id 4.255 -It amrc -Il 532 -If 0 -Ii 86 -Iv 32

Node-4 == Received Ack

r -t 4.973960326 -Hs 6 -Hd 6 -Ni 6 -Nx 440.00 -Ny 90.00 -Nz 0.00 -Ne -1.000000 -Nl RTR -Nw --- -Ma
13a -Md 6 -Ms 5 -Mt 800 -Is 5.255 -Id 6.255 -It amrc -Il 1044 -If 0 -Ii 80 -Iv 29
Node 6:: I Received the Data Packet

Node-6 == Sending Ack back to node 5
d -t 4.973960326 -Hs 6 -Hd 6 -Ni 6 -Nx 440.00 -Ny 90.00 -Nz 0.00 -Ne -1.000000 -Nl RTR -Nw --- -Ma
```

**FIGURE 5.10:** Reconstruction of the original route, see from the trace file.

From the experimental results we obtained, the following graphs (FIGURE 6.1 and FIGURE 6.2) can be drawn which represents the comparison between MRC and EMRC for the time taken to transmit the data from source to destination.

In this graph, X-axis represents the packets that are transmitted in the network and Y-axis represents the time taken in seconds for transmission of each packet. The graph shows that the packets are transmitted using the backup route (long route) in MRC and original route (short route) in EMRC after link/node failure recovery which shows that the performance of the EMRC scheme is better than the MRC scheme.

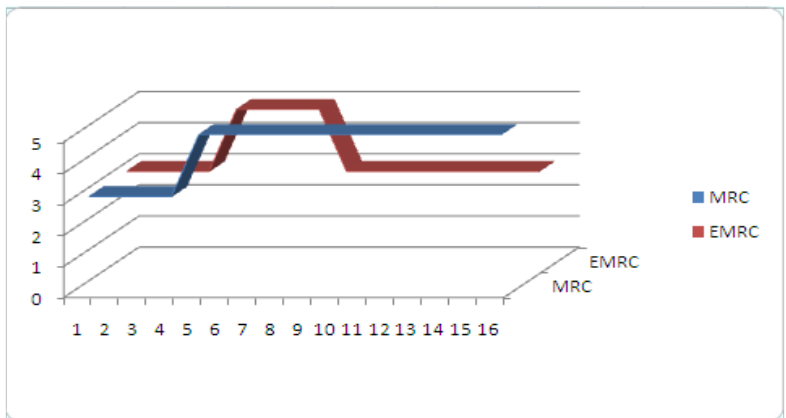

**FIGURE 6.1:** Time taken for transmission of each packet in MRC and EMRC.

In this graph, X-axis represents the number of packets transmitted in the network and Y-axis represents the average time taken for each packet transmission in seconds. The graph shows that the average time taken for each packet transmission in MRC is more than that of in EMRC which shows that the EMRC scheme is more efficient than the MRC scheme.

T Anji Kumar & Dr MHM Krishna Prasad

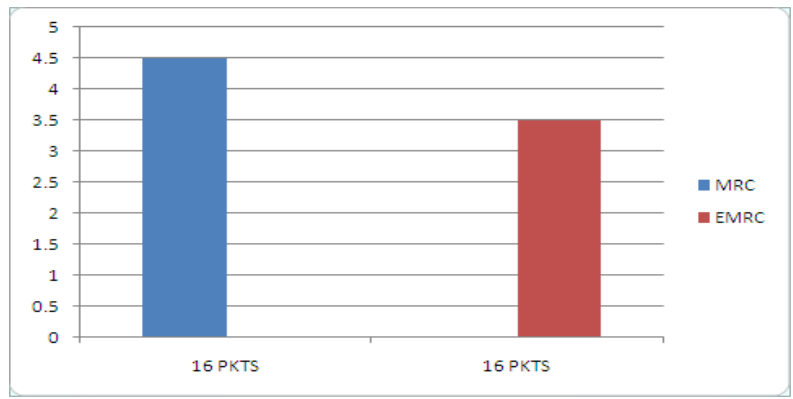

**FIGURE 6.2:** Average time taken for each packet transmission in MRC and EMRC.

## 4. CONCLUSIONS

Multiple Routing Configurations [MRC] recovers network from single node and link failures, but does not support for multiple node/link failures. Enhanced Multiple Routing Configurations [EMRC] is an approach to achieve fast recovery from multiple failures in IP Networks by using the timeslot mechanism. EMRC is based on providing the routers with additional routing information, allowing them to forward packets along routes that avoid a failed component. EMRC guarantees recovery from any failures in source to destination transmission, by calculating the alternate backup configurations in advance.

After the occurrence of original route failure, it is not discarded before completion of timeslot. Within the timeslot, if the failure is recovered, data is transmitted by using the original route. If the failure is not recovered; data is transmitted by using the backup route. During this transmission at any time, if the original route is recovered, data transmission using backup route is stopped and again shifted to the original route. By using this configuration one can improve the fastness of failure recovery and data transmission. EMRC thus achieves fast recovery with a very limited performance penalty.

EMRC does not take any measures towards a good load distribution in the network in the period when traffic is routed on the recovery paths. Existing work on load distribution in connectionless IGP networks has either focused on the failure free case or on finding link weights that work well both in the normal case and when the routing protocol has converged after a single link failure. *Hence, EMRC leaves more room for optimization with respect to load balancing*.

In spite of these encouraging results, this configuration is not to explain some of the issues those are like that this configuration can't develop for some multiple data failures at a time like occurrence of isolated nodes. It is recovered by improving the efficiency of isolated nodes by using the isolated links as restricted links.